\documentstyle[prd,aps]{revtex}

\input psfig.sty
\newcommand{\be}{\begin{equation}}
\newcommand{\ba}{\begin{eqnarray}}
\newcommand{\ee}{\end{equation}}
\newcommand{\ea}{\end{eqnarray}}

\begin{document}
\draft
\title{Topological Pattern Formation}

\author{Janna Levin and Imogen Heard}
\address{Astronomy Centre, University of Sussex}
\address{Brighton BN1 9QJ, UK}
\address{janna@astr.cpes.susx.ac.uk}

\twocolumn[\hsize\textwidth\columnwidth\hsize\csname
           @twocolumnfalse\endcsname

\maketitle
\widetext

\begin{abstract}

We provide an informal discussion of pattern formation in a finite
universe.
The global size and shape of the universe is revealed in the pattern
of hot and cold spots in the cosmic microwave background.
Topological pattern formation can be used to reconstruct the geometry
of space, just as gravitational lensing is used to reconstruct the
geometry of a lens.

\end{abstract}

%\begin{picture}(0,0)
%\put(410,170){}
%\end{picture} \vspace*{-0.15 in}

\bigskip
\medskip

\centerline{Contribution to the conference proceedings for the 
``Cosmological Topology in Paris'' (CTP98) workshop.}

\bigskip
\medskip

]

\narrowtext
%\section{}

We have all come to accept that spacetime is curved.
Yet the idea that space is topologically connected still meets with
resistance.
One is no more exotic than the other.
In the true spirit of Einstein's revolution, 
gravity is a theory of geometry and geometry has two 
facets: curvature {\it and} topology.  

The big bang paradigm forces us to consider the topology of the universe.
As best as we can ascertain, when the universe 
was created both gravity 
and quantum mechanics were at work.  
Any theory which incorporates gravity and quantum mechanics must
assign a topology to the universe.
String theory is currently the most powerful model which naturally
hosts gravity in a unified framework.
It should not be overlooked that in string theory there are six extra 
dimensions all of which must be topologically compact.
In order to create a viable low-energy theory, the internal dimensions 
are finite Calabi-Yau manifolds.
We naturally wonder why
a universe would be created with
six compact dimensions and four infinite ones.
A more equitable beginning might create all spatial dimensions compact
and of comparable size.  Six dynamically squeeze down while the
other
three inflate.  In fact, it is dynamically possible for inflation of 
$3$-space to
be kinetically driven by the contraction of internal dimensions
\cite{jin}.
Whatever mechanism stabilizes the internal dimensions at a small size
would likewise stabilize the external dimensions at an inversely large
size.
Topology need not be at odds with inflation.

Another interesting possibility is that the topology itself naturally
selects the expansion of $3$-dimensions and the contraction of $6$.  The
topology can create boundary contributions to an effective
cosmological constant.  The sign and magnitude of the vacuum energy
depends on the topology and it is 
conceivable that it selects three dimensions for expansion and three
for contraction in a kind of inside/out inflation.  
In the wake of the recent observational evidence
that there is a cosmological constant today, the pursuit of 
these calculations is worthwhile.  Perhaps 
we are still inflating as the vacuum energy tracks the topology
scale.

Our quest to measure the large-scale curvature of the universe may
also produce a measurement of the topology.
(For a review and a collection of papers see \cite{{lum},{volume}}.)
Topological lensing of the cosmic microwave background (CMB) results in 
multiple images of the same points in different directions.
Pattern formation in the universe's hot and cold
spots reveals the global topology
\cite{{lsdsb},{lbbs}}.
Just as with gravitational lensing,
the location, number and distribution of repeated points will allow
the reconstruction of the geometry.
The circles of Ref. \cite{css} are specific 
collections of topologically lensed
points.

\begin{figure}[tbp]
\centerline{{\quad\quad\quad
\psfig{file=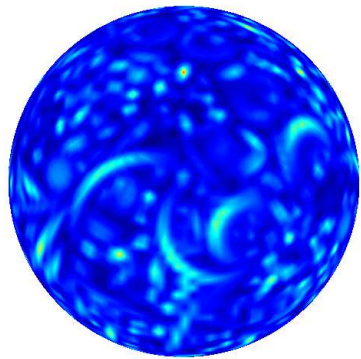,width=2.5in}}} 
\vskip 15truept
\caption{The correlation of every point on the sky with its opposite
in the finite Thurston manifold.
}
\label{thurs_ant}
\end{figure}

We demonstrate topological pattern formation 
%\cite{lsdsb}
with the Thurston space, 
popular in homage to the Thurston person \cite{thurs}.
The space corresponds to $m003(-2,3)$ in the {\it SnapPea}
census \cite{snap}.
A CMB map of the sky does not immediately reveal the geometry.
If we scan the sky for correlations between points we can draw out
the hidden pattern.  There are an infinite number of possible
correlated spheres.  The sphere of fig.\ \ref{thurs_ant} is antipody;
the correlation of  
every point on the sky with its opposite point,
	\be
	A(\hat n)=\left <
	{\delta T(\hat n)\over T}{\delta T(-\hat n)\over T}\right >.
	\ee
In an infinite universe, light originating from opposite directions 
would be totally uncorrelated.  The ensemble
average antipodal correlation would produce a monopole with no
structure.
In a finite universe by contrast, light which is received from
opposite directions may in fact have originated from the same location
and simply took different paths around the finite cosmos.
The antipody map would then show structure as it caught the recurrence
of near or identical sources.  Again, the analogy with gravitational
lensing is apparent.

We estimate antipody following the method of
Ref.\ \cite{lsdsb}.
We take the correlation between two points to be
the correlation they would have in an unconnected, infinite space given their
minimum separation.  
The curvature is everywhere negative and the spectrum of fluctuations
are 
taken to be flat and Gaussian, even in the absence of inflation.  
This is justified on a compact, hyperbolic space
since, according to the tenents of quantum chaos, 
the amplitude of quantum fluctuations are drawn
from a Gaussian random ensemble with a flat spectrum consistent with
random matrix theory.
To find the minimum distance
we move the points under comparison back
into 
the fundamental domain using the generators for the compact manifold.  
The result for the Thurston space with $\Omega_o=0.3$ is shown in 
fig.\ \ref{thurs_ant}.
Notice the interesting arcs of correlated points.  Clearly there is
topological lensing at work.  Arcs were also found under antipody for
the Weeks space in Ref.\ \cite{lsdsb}.
If antipody were a symmetry of the space then at least some circles
of correlated points representing the intersection of copies of the
surface of last scatter with itself
would have been located \cite{css}, as were found for
the
Best space \cite{lsdsb}.
Antipody is by definition symmetric under a rotation by $\pi$ and so 
the back of the sphere is identical to the front.

\begin{figure}[tbp]
\centerline{{\quad\quad\quad
\psfig{file=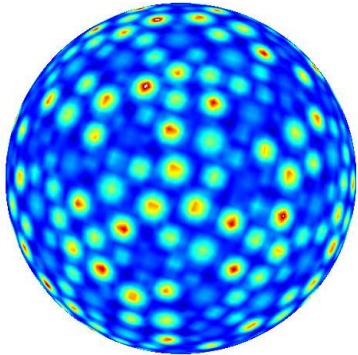,width=2.5in}}} 
\vskip 15truept
\caption{The correlation of one point on the sky with the rest of the
sphere in the Thurston space.  There is a tri-fold symmetry apparent
in the middle of the sphere.
}
\label{origin}
\end{figure}

\begin{figure}[tbp]
\centerline{{\quad\quad\quad
\psfig{file=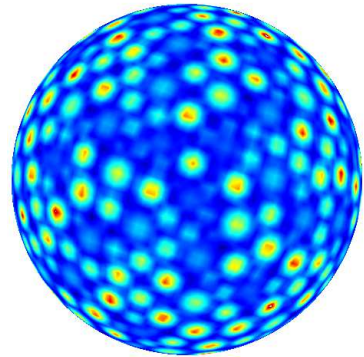,width=2.5in}}} 
\vskip 15truept
\caption{The back of fig.\ \ref{origin}.
The tri-fold symmetry is again apparent
with the three-leaf pattern in the middle of the sphere.
}
\label{origin_back}
\end{figure}

There are an infinite number of correlated spheres which can 
be used to systematically reconstruct the geometry of the fundamental 
domain. Another example is a correlation of one point in the sky with 
the rest of the sphere,
	\be
	C_P(\hat n)=\left
	<{\delta T(\hat n_P)\over T}{\delta T(\hat n)\over T}\right >.
	\ee
This selects out recurrent images of the one 
point.  In an unconnected, infinite space, the sphere would only 
show one spot, namely the
correlation of the point with itself.  In fig.\ \ref{origin} 
we have a kaleidescope of 
images providing detailed information on the underlying space.
There is a trifold symmetry in fig.\ \ref{origin}.  Notice that there
is a band of points moving from the middle upward
vertically which then bends over to the left and that this band repeats twice
making
an overall three-pronged swirl emanating from the middle of the figure.
Since this correlated sphere is not symmetric under $\pi$, we also
show
the back of the sphere in fig.\ \ref{origin_back}.  A different
pattern emerges but still with the tri-fold symmetry.  There is a
three-leaf arrangement of spots in the center of the figure.

We need the improved resolution and signal-to-noise of the future
satellite missions MAP and {\it Planck Surveyor} to observe
topological pattern formation.  High resolution information will be
critical in distinguishing fictitious correlations from real spots.
Beyond the CMB, a finite universe would sculpt the distribution 
of structure on the largest scales.  Even if we never see repeated
images of galaxies or clusters of galaxies, 
the physical distribution of matter
could be shaped by the shape of space.
The topological identifications select discrete modes and the modes
themselves can in turn trace the identifications.  The result is an overall 
web of primordial fluctuations in the gravitational potential 
specific to the finite space.  
A web-like distribution of matter
would then be inherent 
in the initial primordial spectrum \cite{lb}.
This is different from the structureless
distribution of points one would
expect in an infinite cosmos.

We close with the
more fanciful possibility that even time is compact.
If time is compact, every event would repeat precisely as set by the age of
the universe.  
Only a universe which is able to naturally return to its own infancy could be
consistent with a closed time loop.
A big crunch which 
feeds another big bang could allow
our entire history to repeat.   The same galaxies form
and the same stars and planets and people.
Even a proponent of free will can see that at the 
very least we would be limited in the choices we are or are not free to make.
We would live out the same lives, make the same choices, make the same
mistakes.  
Of course, in a quantum creation of the universe,
different galaxies would form in different locations composed of different
stars and new planets.
%You and I 
We would not be here but chances are,
someone would. 
Even if our CMB sky does not look like the Thurston pattern, perhaps
someone's does.

\bigskip
\bigskip

JL thanks the participants and organizers of CTP98.

%\section*{References}

\end{document}